# Insomnia impairs muscle function via regulating protein degradation and muscle clock

Hui Ouyang[1], Hong Jiang[1], Jin Huang[2], Zunjing Liu[1]*

Correspondence: 490644424@qq.com

1 Department of neuromedicine, Peking University People's Hospital,11 Xizhimen South Street, XiCheng District, Beijing 100044, China

2 Yiyang Central Hospital

**Abstract**

Background: Insomnia makes people more physically unable of doing daily duties, which results in a lack of strength, leads to lacking in strength. However, the effects of insomnia on muscle function have not yet been thoroughly investigated. Some researchers also claim that sarcopenic obesity and sleep disturbances are positively correlated. However, the underlying mechanisms of insomnia induced muscle function impairment have not been fully studied yet. So, the objectives of this study were to clarify how insomnia contributes to the decrease of muscular function and to investigate the mechanisms behind this phenomenon.

Methods: We first measured the clinical features and muscle function of 70 moderate or severe insomnia patients. To understand how insomnia influence muscle function, we analyzed the expression level of factors associated with muscle protein degradation (Atrogin-1, FOXO1), muscle protein synthesis (mTOR), protein synthesis/degradation pathways (mTOR, PIK3, FOXO1, Akt) and muscle clock (Per, Bmal). To evaluate statistical significance, data were examined using the Mann-Whitney U test or the two-tailed unpaired Student's t test for two groups. Software programs SPSS 16.0 and GraphPad Prism version 6.0 were used to conduct statistical analyses.

Results: The results showed that lower BMI and grip strength were observed in insomnia patients ($t_{paired}$ = -4.394, $p < 0.001$; $t_{paired}$ = -2.646, $p < 0.01$). The mice in the sleep deprivation(SD) group saw a 7.01 g loss in body mass ($p < 0.001$). The SD group's tibialis anterior and gastrocnemius muscle mass decreased after 96 h of SD ($p < 0.001$). The grip strength reduced in SD group ($p < 0.001$). Using the RT-PCR approaches, we found a significant increase in Atrogin-1, PIK3, FOXO1, mTOR, Akt, Bmal1 and Per expression in SD group versus normal control(NC) group ($p<0.001$).

Conclusions: Insomnia can impair muscle function. The mechanism may be associated with the increased expression of muscle degradation related factors and the activation of PI3K/Akt pathway, as well as the abnormal expression of Clock gene and Per/PI3K/mTOR pathway.

**Introduction**

It is evident that sleep is essential to human life because we spend about one-third of our lifetimes asleep. More people are experiencing sleep deprivation as society continues to shift toward a 24-hour culture[1]. Many individuals find it difficult to get the 7 to 9 hours of sleep per night that are advised due to the increased time demands of modern society[2, 3], a large number of people suffer from insomnia. A recent "public health problem" declaration by the US Centers for Disease Control and Prevention identified inadequate sleep[4]. With the increase of the incidence rate of insomnia related complications, people's interest in sleep debt is also increasing. Lack of sleep every night will reduce physical performance and is associated with poor metabolic health results. Although sleep is important for maintaining health[5], the impact of insomnia is still not fully understood.

According to several epidemiological studies, sleep loss can negatively impact peripheral tissues, including skeletal muscle, and is linked to an increased risk of diabetes, cardiovascular disease, dementia, and all-cause mortality[6-13]. Some authors even reported a positive correlation between sleep disorders and muscular atrophy obesity[14]. As for animal experiments, male Wistar rats lose skeletal muscle mass when they are sleep deprived according to studies[15, 16].

Similar to sleep deprivation, insomniacs have shorter sleep durations and poorer sleep quality. The skeletal muscle system has been shown to be harmed by sleep deprivation (SD). Previous studies have found that long-term sleep deprivation will reduce the muscle strength of athletes. In general, these studies[17-19] show that sleep restriction or prolonged waking time will weaken muscle strength and disturb the physiological response to exercise resistance.

Despite the fact that several research have concentrated on how sleep affects muscles, there are still certain restrictions. The majority of these studies concentrate on the connection between diminished muscle strength and inadequate sleep. The underlying mechanism of muscle function damage caused by insomnia has not been thoroughly studied. In addition, previously published research results are inconsistent. The majority of studies have shown a significant connection between handgrip strength and sleep duration (either short or long or both) [20-23]. However, a few studies found no connection between the length of sleep and hand grip strength[14, 24, 25]. With regard to muscle strength measurement, some studies use self-reported muscle strength, which is not the standard method for muscle strength evaluation. Some studies use the Flinders Fatigue Scale to assess muscle strength [26], which is a good choice for assessing the physical health of the elderly [27]. However, it is still a subjective evaluation method. The evaluation of handgrip strength, which is practical to measure, is the most frequently utilized muscle strength assessment instrument[28] and is highly related to the functioning variables of quality of life[29]. However, it has rarely been used to evaluate the muscle function damage caused by insomnia by researchers.

Although the length of sleep and muscle strength in men are associated, the fundamental mechanisms are still a mystery[21]. As for the mechanism of sleep affecting muscle function, although more and more research is demonstrating the significance of sleep's restorative process for muscular health, the underlying mechanism and specific role of sleep play in muscle function are still unclear. Decreased sleep quality and duration may lead to muscle metabolic imbalance (i.e. between protein synthesis and degradation)[30]. The delicate balance between the rate of protein synthesis and the rate of protein breakdown in muscle controls the amount of skeletal muscle [31]. The muscular mass is preserved when this balance is upheld [31]. Low muscle mass and muscular atrophy are caused by an imbalance between muscle protein production and muscle protein breakdown [32]. The ubiquitin-proteasome system (UPS) is one of the major systems responsible for muscle atrophy. FOXO1 is a part of UPS. As a regulator of MuRF1 and MAFbx expression[33, 34], this E3 ubiquitin ligase can tag myofibrin with ubiquitin, and then to be degraded by 26S proteasome to induce skeletal muscle atrophy[35, 36].

By boosting the expression of muscle-specific ubiquitin ligases, muscular atrophy F-box (MAFbx), and muscle RING finger 1, these proteolytic systems are stimulated (MuRF1)[37-41]. During pathophysiological muscle wasting, ubiquitin ligases family including muscle MuRF1, are supposed to start the breakdown of muscle protein. Under the condition of promoting muscle atrophy, it is reported that muscle RING-finger Protein-1 (MuRF1) and Atrogin-1/Muscle Atrophy F-box (MAFbx), two muscle-specific Ub ligases, both have higher expression levels[42-45], be upregulated under all conditions inducing muscle atrophy examined so far[42-45]. Many variables, including dietary habits, energy levels, mechanical stimulation, pro-inflammatory substances, and oxidative stress, can control the development of muscle atrophy [46-48]. Muscular atrophy may result from sleep deprivation, according to a small number of studies, which may be due to the catabolic process[15, 49, 50]. Consequently, a lack of sleep may control the activation of genes related to muscle atrophy.

The researches hypothesized mechanisms whereby insufficient sleep may disturb baseline muscle protein metabolism, dampen signaling molecules, consequently reducing the growth and protein turnover in skeletal muscle [51, 52]. Animal models could give us fresh perspectives on the joint health issues that might be caused by insufficient sleep [53]. According to a few studies, sleep deprivation can raise rodents' amounts of ubiquitinated proteins[16, 53].

In recent years, the research on the molecular mechanism of biological circadian rhythm has won the Nobel Prize. To investigate the impact of sleep disturbance on muscle, researchers began to pay attention to the role circadian clock plays in muscle. The ability of circadian clocks to not only predict daily variations in the solar cycle but also to changes in the nutrient and oxygen environment reveals a novel mechanism by which peripheral clocks operate that may support rhythmic tissue-specific metabolic fuel selection[54]. One of the tissues with a brisk metabolism, which is identical to that of almost every cell in the body and has circadian rhythms, is skeletal muscle. Previous studies have demonstrated that the control of muscle glycolytic lactate production in myoblasts may be mediated by the circadian cycle [54]. According to the research of Peek et al, the circadian clock gene expression may regulate glucose metabolism and play a crucial part in determining the glycolytic capacity of skeletal muscles. Therefore, the change of muscle clock gene expression may be one of the reasons for the muscle function impairment caused by insomnia.

In general, according to the researches in recent years, sleep disorders may have an impact on muscle function not only by interfering with catabolic and anabolic pathways, but also by affecting muscles' circadian clock gene expression [55]. In response to the sleep-wake and fast-feeding cycles in animals, the circadian clock coordinates the anabolic and catabolic processes in all key organs and tissues; at least in mice, roughly 40% of protein-coding genes exhibit rhythmic transcription [56]. Current literature shows that the disruption of circadian system/clock genes is a risk factor for the development/progression of metabolic syndrome. The BMAL1:CLOCK-targeted expressed genes are greatly concentrated in metabolic, cancer, and insulin signaling pathways in the liver. Several of these genes may play a role in muscle metabolism, transcription, and signal transduction[57].

Thus, we assumed that sleep deprivation may impair skeletal muscle function and causes damage in muscle fibers. Lack of sleep may have an impact on the expression of genes in the muscles' protein synthesis and degradation pathways. As a result, insomnia can upset the equilibrium between the synthesis and breakdown of muscle proteins. Meanwhile, the muscle clock might contribute to the decrease of muscular function brought on by insomnia.

This study's objective is to assess how insomnia affects skeletal muscle and to ascertain the degree to which alterations in protein synthesis and/or protein breakdown contribute to impaired muscular function, and to study the effect of insomnia on muscle clock.

**Methods**

1. Participates

Inclusion criteria: 1) All of the patients were diagnosed with insomnia according to the American Academy of Sleep Medicine International Classification of Sleep Disorders Third Edition 3 (ISCD-3); 2) Age ranges from 18-60 years old; 3) Insomnia Screening Index (ISI) score above 15; 4) Were not working nights or weekends (within the previous three months), 5) had not traveled abroad in the two months prior; 6) Participate in the study voluntarily and sign the informed consent form. From 2021.12.20 to 2022.11.30, 70 moderate or severe insomnia patients were randomly selected from the Department of Neurology of Peking University People's Hospital and Yiyang Central Hospital.

Seventy healthy controls, aged between 18 and 60 years of age, volunteered to participate, were randomly selected to participate in the study. Eligible healthy controls were: 1) did not use any medication; 2) were not putting in shifts (within the previous three months), 3) had not travelled overseas in the past two months; 4) had a regular sleep schedule (6 to 9 hours each night) and no sleep problems that had previously been diagnosed.

2. Muscle function measurement and sleep evaluation of human participates

Using a dynamometer (WCS-100, Nantong, China), hand grip strength was calculated in kg. Standing with their arms normally at their sides, participants clutched the grips as hard as they could. The maximum of twice tests on two hands were adopted. Sleep duration and quality was evaluated by Insomnia severity index (ISI).

3. Other clinical variables

Age, alcohol consumption, cigarette smoking, and body mass index were our covariates (BMI). Cigarette smoking was classed as 'yes' or 'no'. And the consumption of alcohol was rated as "yes" or "no.". By dividing body weight (in kilograms) by height (in meters squared), the BMI was calculated.

4. Mice and generation of sleep deprivation(SD) model

Twelve male C57BL/6 mice (10 weeks old) were divided into the following two groups, each consisting of six mice: (1) the normal control group(NC, n=6), (2) sleep deprivation group(SD, n=6). The animals were kept at a constant temperature of 21 ℃ with a 12-hour cycle of light and darkness (lights on at 7 a.m.). For the entire experiment, water and food were provided. Over the duration of the trial, all animals were housed in one room. For SD group, A multi-platform water environment method that has been adapted was used to create an SD mouse model [57], having been awake for 96 hours straight. The animals (CTL and SD groups) were acclimated to the SD tank for an hour over the course of three days prior to this operation. Every day, between 7:00 and 8:00 am, the animals were handled for a short period of time to measure body mass, clean the SD tanks and housing boxes, and alter feed and water. Only dead or sick mice could not cooperate were taken out of the experiment. The Peking University Health Science Center provided the mice. The animals in this study had an average age of 10 weeks. Following the experiment, all mice were killed via cervical dislocation. All animal experiments were approved by the Animal Care and Use Committee of Peking University People's Hospital.

5. Muscle collection and extraction of tissues

After the trial on day four, the animals were weighed and slaughtered in accordance with the SD procedure by cervical dislocation. Both hind legs' total muscle mass as well as the gastrocnemius and tibialis anterior muscles were dissected, carefully dried on filter paper, weighed, and immediately frozen in liquid nitrogen for biochemical testing. The tissues were kept on ice during the entire process. Fresh muscle samples were thereafter kept at -80 ℃ for biomolecular analysis.

6.Whole muscle lysates

The gastrocnemius muscle was homogenized in a solution containing protease and phosphatase inhibitors along with RIPA buffer (25 mM Tris-HCl, pH 7.6, 150 mM NaCl, 1% NP-40, 1% sodium deoxycholate, and 0.1% SDS) (Thermo Fisher Scientific, US). On ice, the entire surgery was carried out.

7.RNA isolation and quantitative real-time PCR (RT-qPCR)

In order to separate the organic and aqueous phases, chloroform was used to homogenize the gastrocnemius muscle in Trizol reagent (Transgen Biotechnology, Beijing, China). The total RNA was isolated from the aqueous phase by ethanol precipitation. Transgen Biotechnology carried out the RT-qPCR (Beijing). The thermal cycling dice real-time system and SYBR premix ex Taq II were used to measure the expression of Per, Bmal, Atrogin-1, PIK3, FOXO1, Akt, and mTOR (Takara Bio,Shiga, Japan). Every sample was measured twice. To assess any variation brought on by PCR and reverse transcription efficiency, GAPDH was employed as an internal control. The following were the forward and reverse primers used to detect these genes: mus GAPDH-F: 5′-cgagatgggaagttgtca-3′, GAPDH-R: 5′-cgagatggaagcttgtca-3′; mus Per-F: 5′-caattggagcatatcacatccga-3′, Per-R: 5′-cccgaaacacatcccgtttg-3′; mus Bmal-F: 5′-acagtcagattgaaaagaggcg-3′, Bmal-R: 5′-gccatccttagcagcggtgag-3′; mus Atrogin-1-F: 5′-cagcttcgtgagcgacctc-3′, Atrogin-1-R: 5′-ggcagtcgagaagtccagtc-3′; mus PIK3-F: 5′-gaagcacctgaataggcaagtcg-3′, PIK3-R: 5′-gagcatccatgaaatctggtcgc-3′; mus FOXO1-F: 5′-cccaggccggagtttaacc-3′, FOXO1-R: 5′-gttgctcataaagtcggtgct-3′; mus Akt-F: 5′-atgaacgacgtagccattgtg-3′, Akt -R: 5′-ttgtagccataaaggtgccat-3′; 5′-cctcctgaagcaccgttgtg-3′; mus mTOR-F: 5′-cagttcgccagtggactgaag-3′, mTOR -R: 5′-gctggtcataagcgagtagac-3′.

8. SDS-PAGE and western blot analysis

Muscle sample supernatants were collected and kept at -80 ℃ after being lysated. The Beyond TimeTM BCA Test Kit was used to evaluate the protein concentration (Shanghai Beiotim Biotechnology Company, China). SDS-PAGE was utilized for western blotting to separate the same quantity of protein from each sample, which was then transferred to a PVDF membrane (Millipore, USA). The membrane was then treated with the primary antibodies directed against the indicator proteins at 4 ℃ after being incubated with 5% skimmed milk in Tris-buffered saline with Tween20 (TBS-T buffer) for 2 hours. An improved chemiluminescence method was used to identify immune response bands, and Quantity One software was used to analyze the results (Pierce ECL, western blotting substrate; Thermo Fisher Scientific, Pierce, Rockford, IL, US). Each protein band was measured using ImageJ software, and it was then normalized to a loading control (β -actin).

9 Antibodies

The following antibodies were used for Western blotting: rabbit monoclonal anti-Atrogin-1 (ab168372, Abcam), rabbit polyclonal anti-Akt (ab8805; Abcam), anti-PI3K (20584-1-AP, Proteintech), anti-Per1 (13463-1-AP, Proteintech), anti-Bmal (14268-1-AP, Proteintech), anti-mTOR (28273-1-AP, Proteintech), anti-β -actin (T0021, Affinity).

8. Statistical analysis

The Mann Whitney U test or Student's t test were used to assess the data. For data with or without normal distributions, a chi-squared test or a Wilcoxon rank-sum test was employed to compare values of quantitative features between groups. Values are reported as mean ± standard deviation unless otherwise stated, and p<0.05 was regarded as statistically significant. Software GraphPad Prism 6.0 or SPSS 16.0 was used for statistical analysis.

Results

1. Baseline characteristics, muscle function and grip strength of participants

After removing those who were unable to provide the necessary information, a total of 140 participants (63 men and 77 women) qualified for the current study. The median length of baseline sleep for patients with insomnia was 4.4 hours, whereas it was 7 hours for healthy controls. Table 1 displays descriptive statistics for baseline participant traits, muscular function, and grip strength.

In our human investigation, we examined grip strength and muscular function between the insomniac group and the control group. The findings revealed that patients with insomnia had lower BMI and grip strength ($t_{paired}$ = -4.394, p<0.001; $t_{paired}$ = -2.646, p<0.01). These findings supported the findings of our animal investigation and indicated that lack of sleep may affect muscular performance.

Table 1 Basic characteristics, muscle function and grip strength of participants

| Characteristics | Insomnia patients | Normal control | p |
|---|---|---|---|
| Age,year(Mean±SD)a | 44.36±10.24 | 43.34±6.28 | 0.481 |
| Gender,male(%)b | 44.3 | 45.7 | 0.500 |
| Drinking status(%)b | 45.7 | 48.6 | 0.866 |
| Smoking status (%)b | 50 | 50 | 0.99 |
| BMI(Mean±SD)a | 21.76±3.36 | 24.35±3.62 | **<0.001** |
| ISS score a | 24.53±2.16 | 3.51±2.36 | **<0.001** |
| Sleep duration(h) a | 4.43±0.96 | 7.03±1.13 | **<0.001** |
| Hand grip strength,Kg(Mean±SD) a | 31.97±8.69 | 35.93±9.01 | **0.009** |
| Flinders fatigue scale | 18.23±3.73 | 4.20±1.80 | **<0.001** |

Bold for $p \leq 0.05$ based on Student's t-test.

a. Independent-samples T test.

b. Chi-square test for qualitative variable.

2、Muscle mass and grip strength of sleep deprivation(SD ) mice

Table 2 displays the characteristics of the animals. SD mice had a lower average body weight than mice in the control group. The gastrocnemius and tibialis anterior muscles' absolute and combined masses were measured. The SD group's absolute muscle mass was lower than that of the control group. In comparison to the NC group, the wet weight of the gastrocnemius, tibialis anterior, and overall muscles clearly decreased in SD mice. The grip strength of the SD group considerably decreased when compared to the control group ($p<0.001$). (Table 2). The animals in the SD group have so far experienced a 7.01 g decrease in body mass ($p<0.001$). After 96 hours of SD, there was a decrease in the SD group's tibialis anterior muscle mass ($p<0.001$) as well as a decrease in the SD group's gastrocnemius muscle mass in comparison to the NC group ($p<0.001$).

Table 2 Weight of muscles of C57BL/6 mice in SD group and normal control group

|  | Control | SD | *P* value |
|---|---|---|---|
| Tibialis anterior muscle(g) | 0.64±0.03 | 0.45±0.04 | **<0.001** |
| Gastrocnemius muscle(g) | 1.26±0.09 | 0.95±0.09 | **<0.001** |
| Total muscle mass(g) | 4.06±0.46 | 3.32±0.55 | **<0.001** |
| Body weight(g) | 28.84±0.65 | 21.85±0.94 | **<0.001** |
| Grip strength(g) | 206.44±40.40 | 184.11+47.34 | **<0.001** |

Bold for *p*≤0.05 based on Student's t-test.

Data are presented as mean± SEM (n=6); SD, sleep deprivation

3、The expression of muscle atrophy and muscle synthesis markers

We discovered a substantial uptick in Atrogin-1 expression in the SD group compared to the NC group using RT-PCR methods ($p<0.001$). (Fig.1A). In the three days following SD, PIK3 relative expression significantly increased in the SD group compared to the NC group ($p<0.001$) (Fig. 1D), and FOXO1 expression significantly increased in the SD group compared to the NC group ($p<0.001$). (Fig.1B). Akt and mTOR expression were observed to be significantly higher in the SD group compared to the NC group ($p<0.001$). (Fig. 1E,1C).

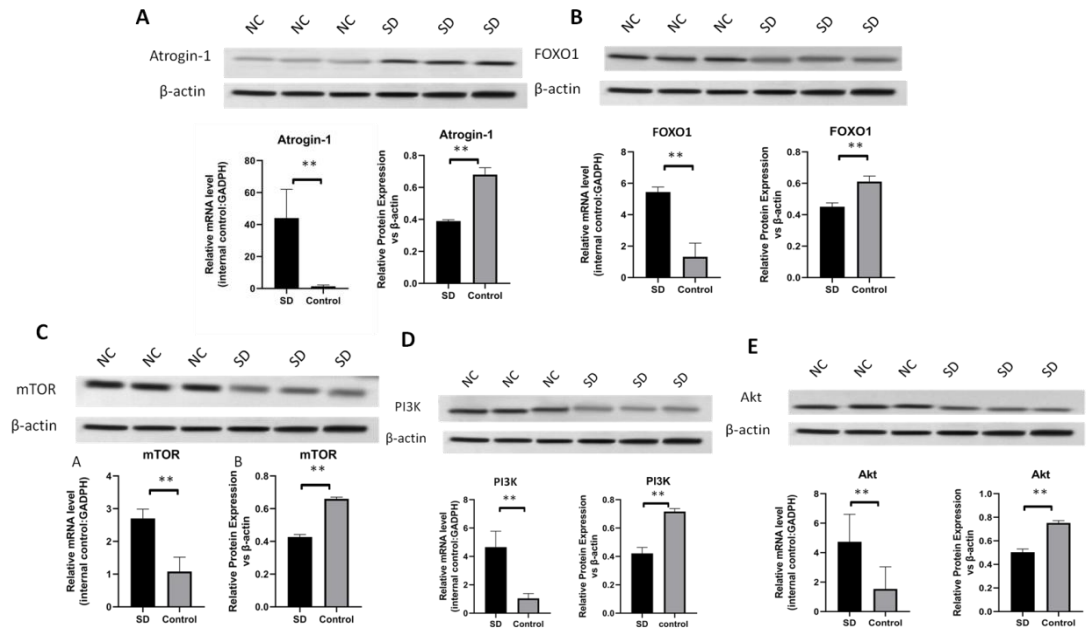

Fig.1 Atrogin-1, PIK3, FOXO1, Akt and mTOR mRNA and protein expression in gastrocnemius muscle of SD and Control mice, where relative mRNA levels and protein are representative of the mean value; n=6 per group.**$p<0.001$.

These findings collectively imply that SD dramatically increases both global protein production and breakdown in skeletal muscle.

4、SD attenuates the expression of Clock genes in murine skeletal muscle

Unexpectedly, circadian core clock gene expression increased significantly (Bmal1 and Per, $p<0.001$) after 96 hours of sleep deprivation (Fig.2). In all, the data shows that SD considerably changes how the circadian core clock genes are expressed in mouse skeletal muscle.

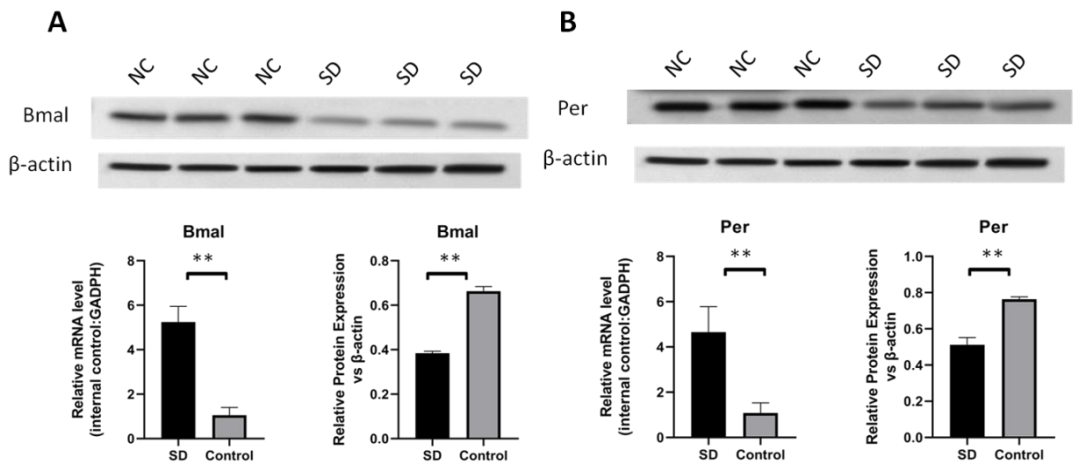

Fig.2 Bmal and Per mRNA and protein expression in gastrocnemius muscle of SD and Control mice, where relative mRNA and protein levels are representative of the mean value; n=6 per

group.**p<0.001.

5、The expression of biological pathways(PI3K/Akt/FOXO,Per/PI3K/mTOR) associated with muscle protein synthesis/degradation

Three days after SD, we discovered a substantial rise in PI3K expression in the SD group compared to the NC group using the RT-PCR and WB methods (p<0.001) (Fig.3). FOXO1 expression considerably increased in the SD group compared to the NC group 96 hours after sleep loss (p<0.001). Akt expression was observed to be significantly higher in the SD group compared to the NC group (p<0.001). mTOR expression was found to be higher in the SD group compared to the NC group (p<0.001). Overall, the findings indicate that, in response to 96 hours of SD, both protein synthesis and degradation pathways (PI3K/Akt/PGC1 and PI3K/Akt/FOXO,Per/PI3K/mTOR) are activated.

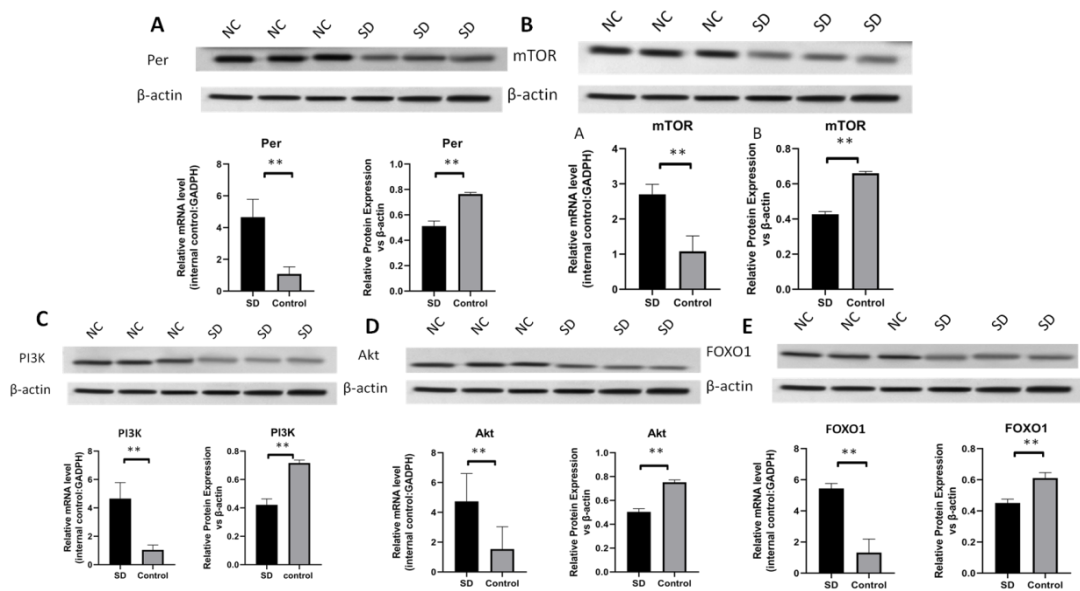

Fig.3 Akt, PIK3,FOXO1 and mTOR mRNA and WB expression in gastrocnemius muscle of SD and Control mice, where relative mRNA and protein levels are representative of the mean value; n=6 per group.**p<0.001.

## Discussion

It is speculated that insomnia may cause muscle function impairment. Current studies, however, have come to conflicting results regarding how sleeplessness affects muscle. Also, the mechanism underlying the muscular atrophy brought on by sleeplessness is still not completely understood. We conducted an extensive study to investigate the affect of sleeplessness on muscles and its underlying mechanisms, utilizing both human samples and animal models. In order to improve the accuracy and reliability of the results, we adopted the following two methods: 1. The patients included in our study were those diagnosed with moderate or severe insomnia in our insomnia clinic, rather than self-reported 'insomnia patients' selected from the community. 2. The patients' muscle function was evaluated not only by subjective scale, but also by grip strength (an objective method). In the experiment, at two levels, we assessed the degree of muscle atrophy: (1) Morphometric, which measures the mass of muscle and (2) Molecular, through the evaluation of protein synthesis and degradation pathways using RT-PCR or western blot techniques.

The main findings of this work were: (1) Moderate and severe insomnia patients have a subjective sense of fatigue and lacking in strength, meanwhile, objective measurements shows that their muscle function and muscle mass are reduced. (2) Sleep deprivation decreases muscle weight and grip strength of mice indicating that sleep deprivation may lead to muscle function impairment. (3) Molecular markers of muscle atrophy and muscle protein degradation (FOXO1 and Atrogin-1) significantly increased after sleep deprivation; (4) The impact of insomnia on the expression of CLOCK genes in muscles may be one of the causes of insomnia-induced muscle function impairment, as the expression of CLOCK genes (Bmal and Per) in muscle increased following sleep deprivation.

According to the study's findings, patients with insomnia had lower muscle strength compared to the healthy control group. Some researchers supposed that these changes may partly due to the metabolic adaptations of SD to the increased metabolic demand[59, 60]. The findings of our study agree with the findings of the majority of earlier investigations. According to some earlier studies, certain insomniacs' muscle function and volume did not significantly decline when compared to the healthy control group [61]. We hypothesize that the primary cause of the disparity between our study's findings and those of several earlier researches may be that the patients we included in the study are all patients who have been definitely diagnosed with moderate or severe insomnia, mild insomnia patients were not included in our study. Some previous studies were conducted on people from the community with self-reported insomnia, or patients with mild insomnia. Hence, different inclusion criteria of research objects may be the reason for the differences in research results.

By maintaining a balance between protein production and protein breakdown, skeletal muscle mass is preserved. As protein synthesis is outpaced by protein degradation, a complex process known as muscle atrophy takes place. Greater decrease of muscle mass is associated with an increase in protein breakdown pathways. Protein degradation is intimately related to the expression of a number of atrophy-associated genes, or "atrogenes," particularly those implicated in the ubiquitin proteasome pathway [62]. The three atrophy-associated genes that are most frequently studied, FOXO1 and Atrogin-1, are linked to protein breakdown. FOXO1 may have a significant role in the control of atrogenes like E3 ubiquitin ligases, particularly Atrogin-1[63, 64]. Protein synthesis is adversely regulated by Atrogin-1. Atrogin-1 may therefore effectively regulate muscle turnover by regulating both protein production and breakdown. While mTOR is associated with protein synthesis, the up-regulation of the ubiquitin proteasome system, which includes E3 ubiquitin ligases, is one of the primary causes of the development of muscular atrophy and muscle wasting. The expression of PI3K and mTOR were additional elements frequently employed to evaluate the activation of protein synthesis.

Current research has confirmed that the muscle function of insomnia patients is impaired and the grip strength is decreased. It has also revealed distinct muscle weight and molecular changes of muscle protein synthesis/degradation pathways after sleep deprivation (SD). Both the indicators of the pathways responsible for myofibrillar protein synthesis and the markers of the pathways responsible for its breakdown have increased, according to our research. The results of our research showed that the main mechanism of insomnia induced muscle function impairment is the increased muscle protein degradation, which is consistent with previous hypothesis that a potentially catabolic environment is generated by SD protocols[15, 49, 65-67]. Different from previous researches[68], we found that muscle synthesis in sleep deprivation mice also increased, which is inconsistent with the previous hypothesis that sleep deprivation leads to reduced muscle protein synthesis. According to the results above, the increase of muscle protein degradation may be the main cause of insomnia related muscle function impairment. The skeletal muscle faces a significant challenge from the sleep deprivation protocol-induced stress due to the increased metabolic requirement [69]. Therefore, increased expression of proteins relates to muscle synthesis after sleep deprivation may be a compensatory response to increased muscle protein degradation, or could partially result from the metabolic adjustments made in response to the increased metabolic demand brought on by lack of sleep and oxidative stress. Hence, the mechanism of sleep deprivation induced muscle atrophy may be different from muscle atrophy caused by other reasons, such as muscle atrophy caused by disuse.

It is well-known that severe muscular atrophy brought on by inactivity is principally caused by a reduction in protein synthesis. Many studies have demonstrated that inactivity causes a considerable reduction in protein production without changing the indicators of protein breakdown [70-72]. Weather increased protein degradation take part in the process remains a topic to debate [73, 74]. Since the markers for protein degradation (Atrogin-1 and FOXO1) are unchanged after disuse and the data indicate that increases in protein breakdown do not necessarily occur in all muscles subjected to unloading, the majority of studies come to the conclusion that loss of muscle mass caused by disuse is caused by reductions in protein synthesis [70, 75]. Therefore, we suspect that the mechanism of disuse-related muscle atrophy may be associated with the insufficient secretion of muscle growth stimulating factors due to reduced activity, which impair muscle protein synthesis. While the mechanism of insomnia induced muscle atrophy is mainly because of abnormal muscle metabolism and the accelerated muscle protein degradation. Therefore, different mechanisms may exist for muscle function impairment caused by different reasons. For example, the treatment of disuse-associated muscle atrophy should mainly use drugs that stimulate muscle protein synthesis, while for insomnia induced muscle function impairment, Strategies to control muscle energy metabolism and lessen muscle protein breakdown can be the focus of potential therapies.

The findings of this study demonstrated that sleep deprivation can change the expression of clock genes in muscle in addition to genes involved in protein synthesis and degradation. A large number of skeletal muscle genes, many of which are involved in transcription, myogenesis, and metabolism, are modulated by circadian rhythms and the molecular clock mechanism [76]. The production and regulation of circadian rhythms depend on clock genes [77-79]. The metabolic processes taking place in skeletal muscle and other peripheral tissues are linked with circadian cycles [80].

Loss of the core clock gene Bmal1 causes severe muscle pathology, including fiber-type shifts, decreased mitochondria, impaired mitochondrial respiration, altered sarcomeric structure, and diminished function, as have previous studies demonstrated that clock genes play a significant role in regulating muscle structure and function[81, 82]. According to studies, skeletal muscle anaerobic glycolysis is regulated by circadian clock-HIF interactions, proving that the choice of oxidative vs glycolytic fuel is really regulated by the clock in skeletal muscle [54].

Few previous studies have explored the effect of insomnia on the expression of clock gene in muscle. Therefore, the results of this study preliminarily revealed the impact of insomnia on the expression of clock gene in muscle, which may provide a new clue to reveal the mechanisms of insomnia induced muscle impairment. In other words, the imbalance of muscle protein synthesis and degradation is not the only mechanism that insomnia affects muscle function; Insomnia may impact muscle function by regulating the expression of muscle clock gene. This may explain why there are still subjective reports of muscle weakness and objective measurements of muscle strength decline in some insomnia patients without muscle atrophy or muscle mass reduction. The mechanism of insomnia affecting muscle function by regulating the expression of clock gene may be an interesting area worth exploring.

With respect to the pathways in muscle influenced by muscle atrophy, we generally found that both pathways in protein breakdown and protein synthesis were activated in muscles of SD mice compared to normal controls. Furthermore, the molecular analyses point to a heightened reaction to SD, with greater activation of degradation pathways in the muscles of mice. Following the sleep deprivation, mRNA transcripts of PI3K/Akt/FOXO, Per/PI3K/mTOR pathways were discovered to have altered expression levels; these changes were mostly caused by increases in muscle protein production and breakdown processes.

A critical intracellular signal transduction pathway is the PI3K/AKT pathway in the process of cell cycle. It is associated with cell stasis, proliferation, and cancer [83]. Translation-related pathways, including the mTOR pathway, are involved in skeletal muscle hypertrophy. Some studies have found that regulation of PI3K/AKT/mTOR signal transduction pathway can affects myocardial cells proliferation [84]. These results suggested that sleep deprivation may lead to muscle protein synthesis/degradation imbalance by simultaneously activating protein synthesis(PI3K/mTOR) and protein breakdown (PI3K/Akt/FOXO) pathways. An increase in expression levels of genes associated with circadian rhythm has also been found. Alterations to core circadian clock genes may be associated with muscle protein synthesis by regulating Per/PI3K/mTOR pathway. One of the reasonable explanations for the mechanisms of insomnia induced muscle function impairment may be that SD leads to the activation of muscle protein synthesis associated pathways. Meanwhile, SD activates the protein synthesis related pathway by increasing the expression of clock gene, which is unable to compensate for muscle atrophy and muscle function damage caused by SD induced muscle protein degradation.

**Limitation**

There are several limitations of this research existed. 1. The sample size for this single-center investigation is relatively limited, which may affect the extrapolation of results to some extent. 2. At present, there is no cell model that can simulate insomnia. Therefore, this study has only conducted research from the clinical and animal experimental levels. Moreover, the specific regulation mechanism of gene expression changes in muscle cannot be further studied because the experiment cannot be conducted on the insomnia cell model. 3. Because few insomnia patients are willing to conduct muscle biopsy, it is difficult to obtain muscle samples of patients for pathological sections and experiments. Therefore, human muscle samples are not used in this study.

**Conclusion**

Insomnia can affect muscle function and even cause muscle atrophy. The PI3K/Akt pathway may be activated and the increased expression of muscle degradation-related factors could be the underlying mechanism. The abnormal expression of clock gene caused by insomnia may also impair muscle function, and the involved pathway may be Per/PI3K/mTOR pathway.

Competing interests

The authors declare that there are no conflicts of interest regarding the publication of this paper.

Ethics approval and consent to participate

The study has been has been approved by the ethics committee of Peking University People's Hospital, China. All participants provided informed written consent to participate in this study. All ethic procedures were performed were in accordance with the ethical standards of the Declaration of Helsinki(World Medical Association 2013).

Consent for publication

Not applicable

Availability of data and materials

Authors declare that data and materials described in the manuscript are freely available to any scientist wishing to use them, without breaching participant confidentiality. The contact should be made via the corresponding author 490644424@qq.com

Funding

This study was funded by Youth's Cultivating Funding of Peking University, Healthscience Center, BMU2022PZYB015 and the medical service and support capacity improvement project from the National Health Commission (Grant No.2199000731)

Acknowledgement

We thank Prof. Jun Zhang and Yongjie Li for their helpful advice.

Authors' contributions

Conceptualization, HO; Methodology, ZL,HO and HJ; Software, HO and HJ; Validation, ZL; Writing-Original Draft preparation, HO and JH. Data collection, HO and JH. All authors have read and approved the final submitted manuscript.